# FOSS-BASED GRID-COMPUTING

A. Mani

Researcher, University of Calcutta

E-mail: a.mani.cms@gmail.com

Home-Page: http://www.logicamani.in

ABSTRACT : In this expository article we will be primarily concerned with core aspects of Grids and Grid computing using free software with some emphasis on utility computing. It is based on a technical report entitled 'Grid-Computing Using GNU/Linux' by the present author.

Grids have made great progress in the area of scientific computing and collaboration projects in the recent past. They have also moved into the domain of business computing more recently. GNU/Linux and Free and Open Source Software (FOSS) on the other hand have steadily progressed into every sphere of computing. The future holds a great lot more for the three. Progress in Grids will all be for GNU/Linux as Grids are made for the *nixes. In this expository article we will be primarily concerned with core aspects of Grids and Grid computing using FOSS with some emphasis on utility computing.

Foster and Kesselman [B2] define a computational Grid "as a hardware and software infrastructure that provides dependable, consistent, pervasive, and inexpensive access to high-end computational capabilities. Grid computing is concerned with coordinated resource sharing and problem solving in dynamic, multi-institutional virtual organizations. The key concept is the ability to negotiate resource-sharing arrangements among a set of participating parties (providers and consumers) and then to use the resulting resource pool for some purpose."

A Grid must

➢ Coordinate decentralized resources through a decentralized shared mechanism. The users and resources will be in different domains and almost all control must be shared.

➢ Use standard, open, general-purpose protocols and interfaces.

➢ Deliver nontrivial quality of services.

A Grid is built from multi-purpose protocols and interfaces that deal with resource discovery, authentication, authorization and resource access in particular. It is necessary that these protocols and interfaces be standard and open, else the system is an application specific system.

All of the above criteria does leave some room for debate on what a grid actually is (see Bhuyya et.al [G1]). But it is accepted that systems like Sun's [Sun Grid Engine], Platform's [Load Sharing Facility], or Veridian's [Portable Batch System] are not grids as they involve centralized control of hosts and have complete control over user requests and allocation. The above three criteria apply most clearly to the various large-scale Grid deployments used within the scientific community.

These include distributed data processing systems like GriPhyN, PPDG, EU DataGrid, iVDGL, DataTAG and the TeraGrid. These systems integrate resources from multiple institutions despite each using their own policies and mechanisms. They use open, general-purpose (Globus Toolkit) protocols to deal with negotiating and managing sharing, security, reliability, and performance.

The next generation of IT evolution is bound to involve 'Utility Computing' in a big way. The design of which is based on a service provisioning model, where users or consumers pay providers for using computing power only when they need to.

The main benefits of the utility computing model for service providers are:

- The computing service provider need not set up actual hardware and software components to satisfy a single solution or user, as in the case of traditional computing.
- Providers can reallocate resources with ease by the use of virtualized resources, that can be created and assigned dynamically to various users when needed.
- The operational costs for providers are reduced due to better resource utilization. The TCO is also reduced.

The design aims and benefits of grids are naturally suited for use as utility computing environments. The interoperability of grids is substantially enhanced by the use of open standard service-based architectures. This makes grids all the more suitable for utility computing. Grid applications have been and are being mostly used in scientific research and collaboration projects. In recent times there has been a great increase in the number of Grid applications in business and industry-related projects too.

Grids provide the following benefits:

- Access to extra resources needed for solving problems that were previously unsolvable due to lack of resources.
- Transparent and instantaneous access to geographically distributed resources of a heterogeneous nature (including hardware and software).
- Improved productivity with reduced processing time.
- The infrastructure for aggregation of resources from multiple sites to meet sudden demands.
- Infrastructure for utilizing under-utilized or unused computing resources that are otherwise wasted.
- Optimal utilization of computing facilities to justify IT capital investments.
- Infrastructure for coordinated resource sharing and problem solving through virtual organizations that facilitate inter departmental and organizational collaboration.
- The gross effort needed for administration is reduced in comparison to managing multiple stand-alone systems.

## GNU/Linux and Grid Computing:

Grid computing by definition must be based over Free and Open Source software and operating systems. Commercial closed source operating systems are hindered in the grid-computing sphere by a wide variety of problems including flexibility, security, integration with applications, lack of tools and scalability among others. That is if we choose to relax the primary criteria. Among the different Free and Open Source operating systems available, GNU/Linux is most suited for the grid computing arena too.

Some of the reasons for this are as follows:

- The Open Grid Services Architecture (OGSA) and the Globus toolkit have been built in compliance with Open standards and free software licenses. The latter deals with the issues of information discovery, security, portability, resource management, data management, communication and error analysis in the Grid context. The development of these have followed the same open line of development of GNU/Linux. The long-term success of grid computing depends on free and open standards, open software, open infrastructure and the development of grid services for business purposes. GNU/Linux is Free and Open Source operating system that has progressed along similar lines and is perfectly compatible with the maintainability of grid computing standards and orientation towards the development of Grid related services.
- Almost all of the computational grid networks developed in the scientific and computing departments of universities and laboratories have been over GNU/Linux or Unix. The available empirical evidence says that GNU/Linux is the best available operating system for grid computing. Few have dared to risk so much of resources on operating systems like Windows NT or Windows XP.
- Many of the benefits of grids for business purposes require empirical confirmation as the associated optimization problems are not sufficiently tractable. This means that such grids would be developed through smaller increments of resources. The open nature of GNU/Linux along with its rock solid stability can sustain such a development scenario even in the face of low capital investments.
- Almost all of the major commercial forays into the grid computing have been GNU/Linux or Unix centric. GNU/Linux is central to IBM's Grid strategy. Sun has released a GNU/Linux specific version of its Grid software (5.3+). Oracle's 10g package is Grid enabled and runs on GNU/Linux.
- The introduction of closed source software into massive Grids is bound to

generate enough distrust and security concerns to the point of inducing massive wastage of resources.

## Grid concepts and components

In this section, we consider different grid concepts and explain associated terminology in detail.

### Types of resources

A grid is a collection of machines, sometimes referred to as "nodes," "resources," "members," "donors," "clients," "hosts," "engines," and many other such terms. They all contribute any combination of resources to the grid as a whole. Some resources may be used by all users of the grid while others may have specific restrictions.

### Computation

Computers in a grid can vary in CPU speed, architecture, software platform, and other associated factors, such as memory, storage, and connectivity. The computation resources of a grid can be used in the following three ways:

- By running an existing application on an available machine on the grid rather than locally
- By running applications specifically capable of parallel computation and
- By running applications that needs to be executed many times, on many different machines in the grid.

The "Scalability" of a grid is a measure of how efficiently the multiple processors on a grid are used. If doubling the number of processors makes an application complete in half the time, then the grid is said to be perfectly scalable.

Scalability is usually expressed as a percentage with respect to this ideal situation and is necessarily application specific.

**Storage**

A data grid is a grid that provides an integrated view of data storage. Each machine on a grid usually provides some temporary or permanent storage for grid use. Grids can be viewed as having pooled RAM. But current systems are hindered by the relatively slow developments in RAM technology.

Primary storage refers to the memory attached to the processor, while storage using hard disk drives or other permanent storage media is understood as secondary storage. Memory attached to a processor usually has very fast access (but insufficiently so for modern processors) but is volatile. It would best be used to cache data or to serve as temporary storage for running applications. Secondary storage in a grid can be used in many interesting ways to increase capacity, performance, sharing, and reliability of data. Mountable networked file systems are used by many grid systems. These include [XFS](), [Network File System (NFS)](), Distributed File System (DFS), [Andrew File System (AFS)]() and General Parallel File System (GPFS). These file systems differ on performance, security and reliability features. The latest version of NFS has an edge over other file systems, but a careful consideration of desired features is essential for a final choice. The volume management feature of these file systems allow for
- Increase of storage by spanning it across multiple machines (with a unifying file system).
- Elimination of maximum file size restrictions and
- A single uniform name space for grid storage.

The last feature makes it easier for users to reference data residing in the grid,

without regard for its exact location. In a similar way, special database software can distribute an assortment of individual databases and files to form a larger, more comprehensive and accessible database.

## Clusters and Grids

If we have a grid with various components being provided through clusters, then it is best to run these clusters (irrespective of their type) using source or binary versions of GNU/Linux distributions specifically meant for clusters. Mature distributions like the [Rocks cluster distribution](#) [17] make cluster management amazingly easy. This distribution is very active with collaboration from many Universities and the industry.

Rocks makes complete Operating System (OS) installation on a node the basic management tool while earlier clustering toolkits have stressed on comparing the configuration of nodes. The main idea behind the Rocks management policy is that it is easier to reinstall the OS uniformly (up to a known configuration) to all the nodes than to try determining the ones that are out of sync. The OS on a cluster node is seen to be as one that can be updated or modified rapidly. This approach scales exceptionally well as the OS can be installed from scratch in a short period of time and upgrades do not interfere with active jobs.

In a cluster different nodes like the compute nodes, the front-end nodes and the file-server nodes require different sets of specialized software. Rocks has a robust mechanism to produce customized distributions (including security patches) that define the complete set of software for a particular node. The Rocks kickstart process is also advanced enough to be fully automatic. The user guide of the distribution provides screen by screen installation and configuration notes too.

There are a host of other packages for dealing with clusters on GNU/Linux. These include Open Mosix [18] and Oscar in particular. Mosix can be used to deal with market-driven cluster grids also.

In contrast to the situation for clusters, there are no similar ready made distribution for grids. This is only to be expected as the hardware standards for grids cannot be expected to be in place beforehand. The base distribution for different grid components can be compiled from different source distributions. Alternatively customized versions of commercially supported GNU/Linux distributions may be installed. Over this suitable grid software must be installed over different components and configured.

## Standards

Standards for Grids are important for both building robust infrastructure and for enabling general-purpose services and tools. The definition of standard "Inter Grid" protocols is one of the critical problems facing the Grid community today. The Global Grid Forum has been very effective in the development of open standards. From a practical view point the Free and Open Source Globus toolkit is a widely used standard by itself. The Open Grid Services Architecture (OGSA) standards are being developed by the Global Grid Forum. These are based over the Globus Toolkit protocols and address emerging new requirements and web services too. Companies such as IBM, Microsoft, Platform, Sun, Avaki, Entropia, and United Devices have all expressed strong support for OGSA.

## Structure of a Grid

Any grid may be viewed as a collection of computing and storage resources accessed by authorized members of participating organizations and/or groups inside the organizations (VO) through dedicated entry points. Resources are managed by specialized configuration and monitor servers and are distributed over sites by web servers. Access rights are granted by certificates issued by the official authorities of the organizations (CA).

Resource management architectures for grids may be centralized, decentralized or hierarchical. Traditional approaches use centralized policies that require complete state information and a common grid management policy (or a decentralized consensus-based policy). The main goal is to optimize measures of system-wide performance. Often it may become too difficult (because of the complexity of grids) to define an acceptable system-wide performance matrix or a common fabric management policy. This is the main problem with the traditional approaches. Hierarchical and decentralized approaches are better suited to grid resource and operational management and especially when they allow for operation of components within fuzzy limits.

In the utility computing scenario, there exist different economic models for managing and regulating resource supply and demand. This applies to both hierarchical and decentralized approaches. The grid resource broker also serves to mediate between producers and consumers. Both producers and consumers can deploy low-level middle-ware systems on the grid for enabling resources. The core middle-ware (on the producer's grid resources) handles resource access authorization for controlling access to authorized users. The user-level middle-ware (on the consumer's machines) allows the consumers to grid-enable applications or produce the necessary coupling for executing legacy applications on the grid. After authentication, consumers interact with resource brokers to execute their application on remote resources. The resource broker takes care of resource discovery, selection, aggregation, data and program translation involved in the process.

Given all this a Grid can look like this from a functional point of view (modified from [G2]):

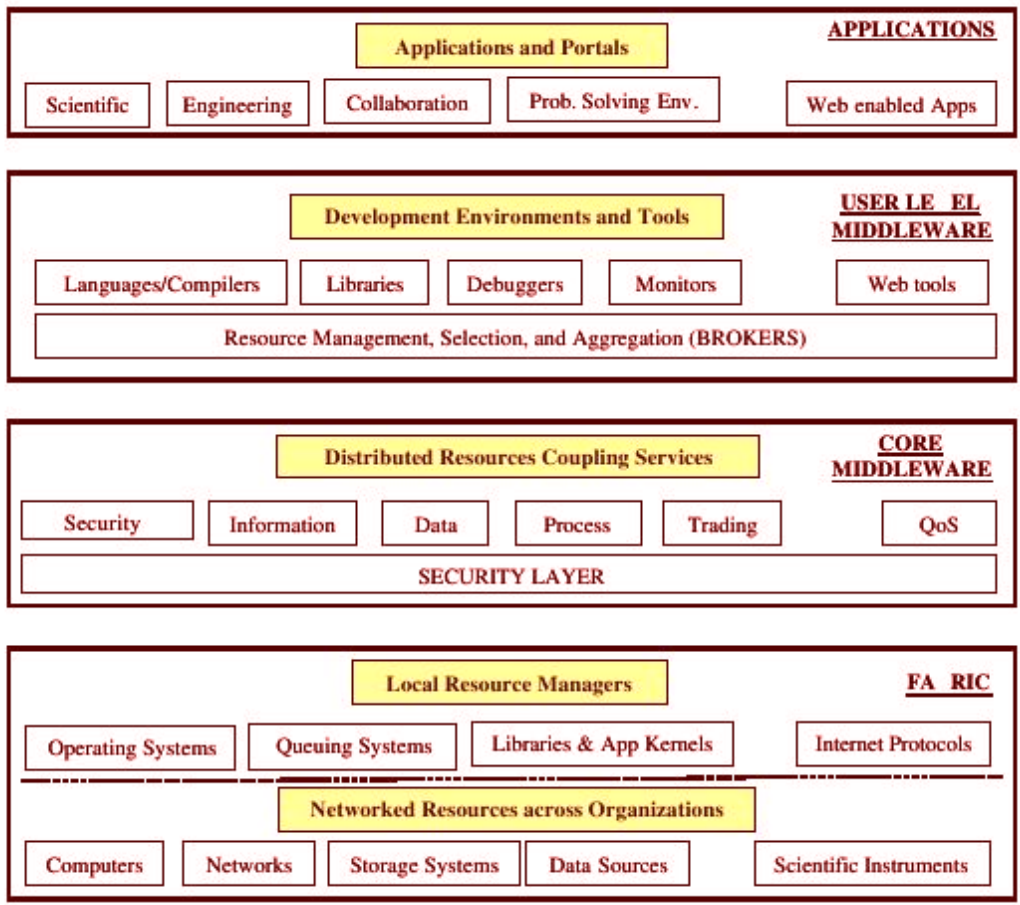

*fig 1: Structure of a Grid*

## Grid Hardware

Grids can be built with a wide variety of hardware and in a variety of configurations. Many optimizations problems will naturally crop up in physically setting up the actual hardware. These optimization problems as we said before are generally very complex and it is best to go in for simplifications of the problem using combinations of heuristics from approximate subproblems and some full problems. For example,

- The solutions of networking topology over different nodes maybe computable to an extent.
- The degradation of performance of different software under different network speeds over a cluster or a simpler grid may be known.
- Some heuristics about the asynchronous nature of the grid may develop from an intended use perspective. In case of grids including clusters at computation nodes it is sensible to keep a maximum amount of memory at the computation nodes. The level of asynchronous nature of grids is better controlled by the resource broker.

The actual hardware configuration necessarily involves studying simulations of possible configurations. A wide variety of simulation tools are available for the purpose (see [G3]). Grid design may also be simplified substantially by restricting applications to a single language like a form of parallel Java (see [C11]).

So in effect the key steps involved in setting up the hardware include (apart from setting bounds on costs, surveying vendor hardware and accessing existing infrastructure):

- The first step must be estimate the required capabilities of the grid. For this an evaluation of the possible software that would be used is essential.
- If a grid across the existing hardware is computed to be inadequate or otherwise problematic due to bottlenecks and network congestion, only then must a redesign attempted.
- If the required grid must be most capable, then the associated optimization problem must be formulated in mathematical terms and solved. Typical optimization problems relate to congestion minimization, choice of network topologies and coherent handling of storage apart from feasibility problems relating to processor utilization.

- Maximization of the speed and capacity of the network is desirable only when it is justified. The relevant hardware that may be used are indicated in the section on hardware.
- Grid hardware must be added incrementally with possible revision at each stage. This helps in keeping down costs. These steps must be aided by a suitable simulation software ([G5] is compulsory reading).
- The software provided by different bodies and vendors are aimed at substantially simplifying the operation of a grid. Optimization relating to resource allocation, scheduling processes, security are all handled by the better ones. So these will not be a problem if good simulators are used.

The key components that will determine the grid hardware include all of the following:

- Desktop PCs, servers and other computers of different or similar configuration.
- Networking Hardware including some of the following: Gigabit class Ethernet cards or Grid specific Ethernet cards (like the T110 card at 10-Gbit/sec with a latency of 10 microseconds), high speed switches, 10BaseF fiber optic connectors or other high speed cables.
- Clusters of PCs if desired.
- Specialized computer aided instruments or sensors and other computer peripherals (as is desired).
- Additional memory modules for distribution among the grid components.
- Other tools and Simulation Software ([G5]).

From the user's point of view however the grid hardware components include:

- A workstation connected to the network accessing a grid web portal

- a user interface grid entry point (UI) and
- a computing and storage environment.

A typical Grid architecture consists of four layers:
- Fabric layer
- Core middle-ware
- User-level middle-ware and
- Applications and portals layers.

The Grid fabric layer consists of distributed resources that include computers, storage devices, networks, scientific instruments and computer aided equipment. The computational resources consist of clusters, supercomputers, servers and desktops of various types (possibly). Real-time data from satellite based remote sensing networks, telescopes and other sensor networks can be stored in databases and accessed as in different desired ways.

The purpose of Core Grid middle-ware is to offer different services that include storage access, allocation of resources, remote process management, security, information registration and discovery and providing Quality of service like resource trading. These services provide a uniform method for accessing distributed resources. The inherent complexity and heterogeneity of the fabric level is masked in the process.

User-level Grid middle-ware on the other hand uses the above core-level middle-ware interfaces to provide services over those. The services provided include resource brokers for the management of resources and scheduling of allocations, application development environments and programming tools like libraries and simulators.

The last layer is built upon the interfaces and resource brokering and scheduling services provided by the user-level middle-ware. In weather prediction for example the software would need access to computational power, remote and will also need to interact with other CAI. Grid portals provide user interfaces for the end-users through Web-enabled application services.

We describe some of the main components of a grid resource broker in more detail in what follows:

**Grid Resource Broker:**

The Grid resource broker can often be seen consist of the Job Control Agent (JCA), Grid Explorer (GE), Schedule Advisor (SA), the Trade Manager (TM) and the Deployment Agent.

The Job Control Agent is required to ensure the persistence of jobs through coordination with Schedule Advisor for schedule generation, handling actual creation of jobs, maintaining job status and interacting with users, Schedule Advisor and Deployment Agent.

The Grid Explorer interacts with the Grid Information Service (GIIS) for the discovery and identification of resources and their status.

The Schedule Advisor discovers Grid resources using the Grid Explorer. It also selects Grid resources and assign jobs to them (schedule generation) based on user requirements.

The Trade Manager is for accessing the market directory services for service negotiation and trading with GSP(s) on the basis of the resource selection algorithm of Schedule Advisor.

The Deployment Agent activates task execution on the selected resource according to the Schedule Advisor's instruction and periodically updates the status of task execution to job control agent.

**An Example Scenario (Using the Globus Toolkit):**

The following is based on a grid constructed using an earlier version (~2002) Globus toolkit [1]. We describe only some of the essential features for illustrative purposes.

From a physical perspective a grid is composed of a set of sites distributed across many places. Each site in turn is composed of many farms (fabric grid components) and servers. Inside the farms we can have different computers and other hardware. If the grid in question is to be smaller in that it is distributed over a single physical place (as in some smaller organizational grids) then also the following considerations will apply with some alterations. The computing environment is made of a variable number of interconnected nodes belonging to one of the following hardware elements (some integration of components by virtualization is possible):

A configuration node (called the LCFG server), that stores the software packages and tools for installation and maintaining for all local components. One LCFG server is required for each site.

A node (called the Computing Element (CE)) for job distribution to the local computing queues (gatekeeper). Each farm must have a computing element.

A Working Node (WN) actually performs the computation and as many as is permitted by site size, cost and CE supporting capability must be used. A working node may be include PCs, mainframes or clusters.

A user interface (UI) for grid access hosting user accounts. Each site must have one UI.

A Storage Element (SE) node with storage capacity for data archiving and management. The exact nature of this depends on the file system being used too.

SE information is accessed by the Replica Catalog (RC). RC is a hierarchical set of facilities to store node informations like operating system, application support, CPU type and available RAM collected by the Information Index (II). At least one RC must be used per grid.

Each grid must possess a Resource Broker (RB) node. This is for performing job distribution according to resource needs and availability. Often a node for logging and bookkeeping (LB) is integrated into the RB.

**WP1 Tools – Workload and Job Submission**

Job submission is handled by several cooperating tasks that are implemented inside WP1 that is responsible for workload management. The different WP1 components are:

- UI, User Interface
- RB, Resource Broker
- JSS, Job Submission Service

- II, Information Index
- LB, Logging and Bookkeeping

A job must be described to the grid by a text file in a syntax like the JDL syntax that declares the job attributes including the executable file name, input and output files, data access to SE, CPU type, queue manager, memory, etc. Input and output are transmitted using a file caching area called the sandbox.

The job requirements are first sent to the resource broker. The RB then queries the Information Service (IS) and the Replica Catalog (RC) to choose the appropriate CE for job submission. The interaction of RB, JSS and CE results in the generation of a job queue and its execution on one of the worker nodes (WN) members of the chosen CE.

The CE node holds basic information for job handling and performs the gatekeeper tasks that include:
- Acting as a Front-end to the local farm
- Validation of user rights based on the grid map files
- Executing the Job Manager (JM) process and
- Interacts with the local queue to submit jobs to the most suitable WN.

The grid resources are stored in GRIS form at CE level. At the site level they are stored in GIIS form. The II index stores the GIIS and possibly the GRIS IP addresses. This is used for handling the grid resources globally.

User access is achieved by user accounts stored at UI level in the usual Unix account format and at grid level through the grid map-files stored at /opt/globus/etc/grid-mapfile for example. Grid mapfiles map a real or virtual user to a user certificate.

WP1 also includes

- Relational monitoring of R-GMA in collaboration with WP3
- Tools for interactive usage of the grid
- Support for complex jobs executing in cooperation with other jobs
- Grid accounting
- Support for advance reservation of grid resources
- GUI (third party)
- Tools for job partitioning and checking
- job query and logging

**WP2 Tools – Data Storage**

The WP2 tools deal with data storage and handling. In particular they also provide:

- Services for mapping logical file names (LFN) into physical ones (PFN) (this can be done with the help of ldap interface)
- GDMP data mirroring
- Interaction with the BrokerInfo file.

**WP4 Tools – Farm Setup**

WP4 tools deal with the setup and management of farms. The fabric tasks are:

- installation and configuration
- monitoring of sanity and fault tolerance
- configuration management and update
- resource management, like farm status, queue management, etc.

In many grids farm setup is handled by LCFG a client-server tool that performs node clustering. In any site with grid resources, even minimal as a single UI, there should be a node acting as LCFG server. The server stores all tools required to install and maintain nodes of the farming. Each node is installed from the server via a startup image (network image).

The startup image boots the node and starts GNU/Linux installation from the LCFG server. The procedure is similar to standard installation (though with a more granular configuration). For each node the LCFG server must store a configuration table so that the node is installed as CE, SE, UI, WN as planned in the local site. LCFG is much more than an installation tool as all operations on the nodes are controlled and started from the LCFG sever. Items that are configurable and variable are called objects. These are defined and maintained on the LCFG server for each node. User accounting for example is an object and new users are created on each node from the LCFG server.

Software packages are maintained in the same way on the server. Updates can be done as Cron jobs. A very important feature of LCFG is the cleanup of non-LCFG actions at node reboot. To access the grid, users must have valid certificates. Certificates are typically created according to the X.509 standard and can be managed by OpenSSL.

The whole scenario can be seen to fit into the scheme of things indicated in the following figure (Globus manual [1]):

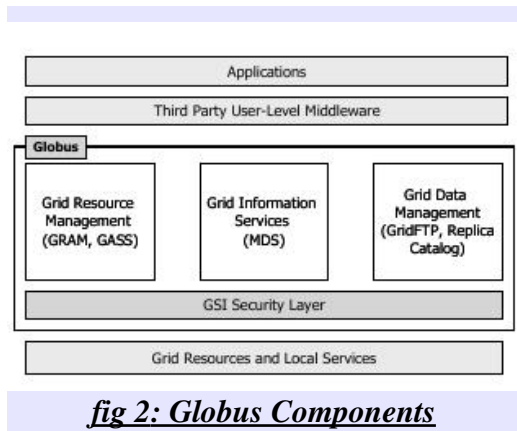
*fig 2: Globus Components*

## Data Grids

The adoption and success of Grids as a utility computing platform depends crucially on the infrastructure that it provides for data management. This means that Grids must be able to effectively store, process, catalog and share data. The associated infrastructure is also referred to as Data Grids. A Data Grid provides services that allow users to discover, transfer and maintain large repositories of data. At the very minimum, a Data Grid provides a high-performance and reliable data transfer mechanism and a data replica management infrastructure. Data manipulation operations in a Data Grid are mediated through a security layer that, in addition to the facilities provided by the general Grid security services, also provides specific operations such as managing access permissions and encrypted data transfers.

A subset of OGSA deals with data services that describe data and the mechanisms to access it. Data services

- Provide for virtual concepts and views of data through multiple views (that are differentiated by attributes and operations).
- Enable different data sources including legacy databases and data

- repositories to be treated in the same way through standard mechanisms.
- Enable applications like data-intensive work-flows through virtual views of data.

An user can either check whether a given virtual data is valid in a relevant sense or reuse it in further computations. The accessibility of virtual data creates many new possibilities in knowledge discovery and dynamic application development. Therefore the QOS parameters associated with a data service include access permissions, data size, permission to modify, relevance and available bandwidth to storage locations.

There are many Grid projects that represent data in terms of computational procedures. An example is the Virtual Data Grid (VDG) project. Projects like Pegasus (a work-flow management system from the GriPhyN project) use VDG to reduce work-flows.

## **GRIA**

GRIA is the first and among the most advanced Grid middle-ware designed specifically for business use across organizational boundaries. The GRIA software is an Free and Open Source Product that supports a wide range of business models for providing and exploiting remote services using standard protocols. It is being used in various organizational grids.

GRIA uses standard Web Service protocols and security mechanisms, and can inter-operate with MONO, .NET applications and services (even though the GRIA packages and API themselves are implemented using Java). GRIA comes with easy-to-use installers and full user documentation, user tutorials and sample applications.

GRIA is supported freely and commercially too.

From an application point of view GRIA allows the integration of legacy computing, cluster facilities, applications and new services within a procurement and billing process. GRIA has a well-defined interface for separating service usage from service management. This applies for a very wide spectrum of applications.

GRIA includes a basic application services package that allows an organization with cluster computing facilities to provide data storage and processing (using applications installed on the cluster) for trusted users too. Other application services can be developed, and some additional packages are also available from the GRIA website.

The key Grid/Web Service standards and specifications adhered to by GRIA include the OGSA and the following (as indicated in the the GRIA manual [12]):

- WS-I Basic Profile and WS-I Basic Security Profile that describe profiles on industry Web Service specifications that promote interoperability.
- WS-Security, a set of SOAP extensions to provide message-level integrity, confidentiality and authentication.
- WS-Federation, which describes how to use WS-Trust, WS-Security and WS-Policy together to provide federation between security domains.
- WS-Addressing describes the encapsulation and use of a (possibly contextual) Web Service address via End Point References (EPR)
- Web Service Resource Framework (WSRF), which describes a particular use of WS-Addressing to access resources via contextual Web Services.
- WS-Notification, which defines a collection of interfaces for transmitting notification messages directly between a producer and a consumer using push-or pull-style transfer, plus a specification for distribution of these messages through a broker, and for defining topics that allow subscribers to select particular notifications of interest.
- SAML, an XML standard for exchanging authentication and authorization data between security domains.

GRIA uses as little of the less well established functionality in the less mature standards and intends to increase conformance in accordance to future developments.

GRIA is very easy to setup and administer over any organizational grid. For more details the user manual may be consulted.

## **Other Grid Software :**

A very large number of Grid softwares is available for performing different Grid related functions. These fall in the categories of middle-ware, front-ends, portals, market-driven Grid-ware, simulators and user-level middle-ware. Some evaluations of these are available in the literature. A detailed study of these is beyond the scope of the present article. The Nimrod-G and Gridbus middle-ware have certain interesting features.

The Nimrod-G resource broker and Gridbus Grid service broker are examples of service-oriented computational Grid brokers for the so-called parameter sweep applications. Job scheduling is done on the basis of economic heuristics and user-defined quality of service criteria(usually a time schedule). The adaptive algorithms used in Nimrod-G (for scheduling parameter sweep applications) are for:
- Cost Optimization: The key criteria is that execution time must be within the specified time schedule and execution cost must be a minimum.

- Time Optimization: The key criteria is that execution time must be a minimum subject to a bounded budget.
- Cost-Time Optimization: Similar to cost optimization, but if there are multiple resources with the same cost, the algorithm tries to apply time optimization for minimizing execution time.
- Conservative Time Optimization: Is cost-time optimization, but the algorithm sees to it that each unprocessed job in the application has a minimum budget-per-job.

The [Gridbus](#) broker extends cost and time optimization to schedule distributed data-intensive applications. These applications require access to and processing of large and storage-wise distributed datasets.

Various commercial vendors provide industrial solutions to support utility computing. Three major industrial solutions for utility computing include [Adaptive](#) Enterprise (HP), [E-Business on Demand](#) (IBM) and [Sun Grid](#) (Sun). All three solutions use Grids as the core enabling technology, though their marketing terminology differ substantially. The following tables from [G3], provide key information relating to some of the market-driven Grid-ware.

TABLE I: Summary of market-based resource management systems

| Computing Platform | Market-based RMS | Economic Model | Brief Description |
| --- | --- | --- | --- |
| Clusters | Cluster-On-Demand * [20] | tendering/ contract-net | each cluster manager uses a heuristic to measure and balance the future risk of profit lost for accepting a job later against profit gained for accepting the job now. |
| | Enhanced MOSIX * [19] | commodity market | it uses process migration to minimize the overall execution cost of machines in the cluster. |
| | Libra * [9] | commodity market | it provides incentives to encourage users to submit job requests with longer deadlines. |
| | REXEC * [7] | bid-based proportional resource sharing | it allocates resources proportionally to competing jobs based on their users' valuation. |
| | Utility Data Center [52] | auction | it compares two extreme auction-based resource allocation mechanisms: a globally optimal assignment market mechanism with a sub-optimal simple market mechanism. |

TABLE I: Continued.

| Computing Platform | Market-based RMS | Economic Model | Brief Description |
|---|---|---|---|
| Agents | D'Agents [17] | bid-based proportional resource sharing | the server assigns resources by computing the clearing price based on the aggregate demand function of all its incoming agents. |
|  | Preist et al. [18] | auction | an agent participates in mutiple auctions selling the same goods in order to secure the lowest bid possible to acquire suitable number of goods for a buyer. |
|  | WALRAS [16] | auction | consumer and producer agents submit their demand and supply curves respectively for a good and the equilibrium price is determined through an iterative auctioning process. |
| Distributed Databases | Anastasiadi et al. [53] | posted price | it examines the scenario of load balancing economy where servers advertise prices at a bulletin board and transaction requests are routed based on three different routing algorithms that focuses on expected completion time and required network bandwidth. |
|  | Mariposa [6] | tendering/ contract-net | it completes a query within its user-defined budget by contracting portions of the query to various processing sites for execution. |
| Grids | Bellagio [54] | auction | a centralized auctioneer computes bid values based on number of requested resources and their required durations, before clearing the auctions at fixed time periods by allocating to higher bid values first. |
|  | CATNET [55] | bargaining | each client uses a subjective market price (computing using price quotes consolidated from available servers) to negotiate until a server quotes an acceptable price. |
|  | Faucets * [21] | tendering/ contract-net | users specify QoS contracts for adaptive parallel jobs and Grid resources compete for jobs via bidding. |
|  | G-commerce [56] | commodity market, auction | it compares resource allocation using either commodity market or auction strategy based on four criteria: price stability, market equilibrium, consumer efficiency, and producer efficiency. |
|  | Gridbus [15] | commodity market | it considers the data access and transfer costs for data-oriented applications when allocating resources based on time or cost optimization. |
|  | Gridmarket [57] | auction | it examines resource allocation using double auction where consumers set ceiling prices and sellers set floor prices. |
|  | Grosu and Das [58] | auction | it studies resource allocation using first-price, vickrey and double auctions. |
|  | Maheswaran et al. [59] | auction | it investigates resource allocation based on two "co-bid" approaches that aggregate similar resources: first or no preference approaches. |
|  | Nimrod/G * [8] | commodity market | it allocates resources to task farming applications using either time or cost optimization with deadline and budget constrained algorithms. |

TABLE I: Continued.

| Computing Platform | Market-based RMS | Economic Model | Brief Description |
|---|---|---|---|
| | OCEAN [60] | bargaining, tendering/ contract-net | it first discovers potential sellers by announcing a buyer's trade proposal and then allows the buyer to determine the best seller by using two possible negotiation mechanisms: yes/no and static bargain. |
| | Tycoon * [61] | auction | it allocates resources using "auction share" that estimates proportional share with consideration for latency-sensitive and risk-averse applications. |
| Parallel and Distributed Systems | Agoric Systems [62] | auction | it employs the "escalator" algorithm where users submit bids that escalates over time based on a rate and the server uses vickrey auction at fixed intervals to award resources to the highest bidder who is then charged with the second-highest bid. |
| | Dynasty [63] | commodity market | it uses a hierarchical-based brokering system where each request is distributed up the hierarchy until the accumulated brokerage cost is limited by the budget of the user. |
| | Enterprise [64] | tendering/ contract-net | clients broadcast a request for bids with task description and select the best bid which is the shortest estimated completion time given by available servers. |
| | Ferguson et al. [65] | posted price, auction | it examines how first-price and dutch auctions can support a load balancing economy where each server host its independent auction and users decide which auction to participate based on last clearing prices advertised in bulletin boards. |
| | Kurose and Simha [66] | bid-based proportional resource sharing | it uses a resource-directed approach where the current allocation of a resource is readjusted proportionally according to the marginal values computed by every agent using that resource to reflect the outstanding quantity of resource needed. |
| | MarketNet [67] | posted price | it advertises resource request and offer prices on a bulletin board and uses currency flow to restrict resource usage so that potential intrusion attacks into the information systems are controlled and damages caused are kept to the minimum. |
| | Spawn [5] | auction | it sub-divides each tree-based concurrent program into nodes (sub-programs) which then hold vickrey auction independently to obtain resources. |
| | Stoica et al. [68] | auction | the job with the highest bid starts execution instantly if the required number of resources are available; else it is scheduled to wait for more resources to be available and has to pay for holding on to currently available resources. |
| Peer-to-Peer | Stanford Peers * [69] | auction, bartering | it uses data trading to create a replication network of digital archives where a winning remote site offers the lowest bid for free space on the local site in exchange for the amount of free space requested by the local site on the remote site. |
| World Wide Web | Java Market [70] | commodity market | it uses a cost-benefit framework to host an internet-wide computational market where producers (machines) are paid for executing consumers' jobs (Java programs) as Java applets in their web browsers. |

TABLE I: Continued.

| Computing Platform | Market-based RMS | Economic Model | Brief Description |
|---|---|---|---|
| | JaWS [28] | auction | it uses double auction to award a lease contract between a client and a host that contains the following information: agreed price, lease duration, compensation, performance statistics vector, and abort ratio. |
| | POPCORN [27] | auction | each buyer (parallel programs written using POPCORN paradigm) submits a price bid and the winner is determined through one of three implemented auction mechanisms: vickrey, double, and clearinghouse double auctions. |
| | SuperWeb [26] | commodity market | potential hosts register with client brokers and receive payments for executing Java codes depending on the QoS provided. |
| | Xenoservers [71] | commodity market | it supports accounted execution of untrusted programs such as Java over the web where resources utilized by the programs are accounted and charged to the users. |

TABLE II: Survey using market model taxonomy

| Market-based RMS | Economic Model | Participant Focus | Trading Environment | QoS Attributes |
|---|---|---|---|---|
| Cluster-On-Demand | tendering/ contract-net | producer | competitive | cost |
| Enhanced MOSIX | commodity market | producer | cooperative | cost |
| Libra | commodity market | consumer | cooperative | time, cost |
| REXEC | bid-based proportional resource sharing | consumer | competitive | cost |
| Faucets | tendering/ contract-net | producer | competitive | time, cost |
| Nimrod/G | commodity market | consumer | competitive | time, cost |
| Tycoon | auction | consumer | competitive | time, cost |
| Stanford Peers | auction, bartering | consumer, producer | cooperative | cost |

TABLE III: Survey using resource model taxonomy

| Market-based RMS | Management Control | Resource Composition | Execution Service | Execution Support | Accounting Mechanism |
|---|---|---|---|---|---|
| Cluster-On-Demand | decentralized | NA | NA | NA | decentralized |
| Enhanced MOSIX | decentralized | heterogeneous | dedicated | time-shared | decentralized |

TABLE III: Continued.

| Market-based RMS | Management Control | Resource Composition | Execution Service | Execution Support | Accounting Mechanism |
|---|---|---|---|---|---|
| Libra | centralized | heterogeneous | dedicated | time-shared | centralized |
| REXEC | decentralized | NA | non-dedicated | time-shared | centralized |
| Faucets | centralized | NA | NA | time-shared | centralized |
| Nimrod/G | decentralized | heterogeneous | non-dedicated | NA | decentralized |
| Tycoon | decentralized | heterogeneous | dedicated | time-shared | decentralized |
| Stanford Peers | decentralized | NA | dedicated | NA | NA |

TABLE IV: Survey using job model taxonomy

| Market-based RMS | Job Execution | Job Dependency | Job Composition | QoS Specification | QoS Update |
|---|---|---|---|---|---|
| Cluster-On-Demand | sequential | NA | single-task | rate-based | static |
| Enhanced MOSIX | parallel | NA | NA | NA | NA |
| Libra | sequential | NA | single-task | constraint-based | static |
| REXEC | parallel, sequential | NA | single-task | constraint-based | static |
| Faucets | parallel | NA | NA | constraint-based | static |
| Nimrod/G | sequential | NA | mutiple-task | optimization-based | static |
| Tycoon | NA | NA | NA | constraint-based | static |
| Stanford Peers | NA | NA | NA | NA | NA |

TABLE V: Survey using resource allocation model taxonomy

| Market-based RMS | Resource Allocation Domain | Resource Allocation Update | QoS Support |
|---|---|---|---|
| Cluster-On-Demand | external | non-adaptive | soft |
| Enhanced MOSIX | internal | adaptive | NA |
| Libra | internal | adaptive | hard |
| REXEC | internal | adaptive | hard |
| Faucets | internal | adaptive | soft |
| Nimrod/G | external | adaptive | soft |
| Tycoon | internal | adaptive | soft |

# Compilers for Grids:

A non-parallel or serial compiler is a program that converts programs written in high-level languages to machine-level instructions. An interpreter simulates program execution for programs written in a source language and necessarily executes instructions one by one. Compilers for uniprocessor computers consist of a front-end that analyses and translate the source program into intermediate representation and a back end that translate intermediate representation to machine instructions. The code is always optimized for minimum execution time and memory consumption. Compilers can be split into a lexical analyzer, syntax analyzer, a parser, intermediate code generators, code optimizers, code generators, symbol table managers and error handlers. These roughly work on a program in the presented order.

In general parallelism can be used to considerably speed up serial programs. The development of parallel code can be done from scratch or by using techniques like vectorization and dependency graph analysis to extract segments admitting of parallel representation in a piece of code written for a serial compiler. Parallel computing is naturally relevant in the grid computing context for speeding up computation as computationally intensive problems can be divided into smaller problems and distributed among independent processors with shared or independent resources. In the grid context scope for dealing with some dynamic scheduling is necessary. This reflects on compilers as a robustness requirement.

# Parallel Programming Interfaces

Different hardware architectures have led to fundamentally different ways of parallel programming. General parallel computers are of two types. In one class of parallel architectures a single unified address space is available for each processor. In these shared memory computers which includes Grids the memory is physically distributed across the system but each processor (if otherwise permitted) is able to access any part of it through a single address space. The Grid hardware is responsible for presenting this abstraction to each processor. Communication between processors is done implicitly through normal memory load and store operations. But there are many finer aspects to this classification as indicated in earlier sections.

In the second class of parallel architectures, message passing is the primary means of communication. These computers usually have separate memory spaces (as in a cluster), are often made up of many individual single processor computers. Messages are sent from processor to processor through the network using software primitives. These two architectures have led to two very different programming interfaces.

The first class of architectures have led to programming interfaces like OpenMP, while the second class to interfaces like [MPI](#) ([Message Passing Interfaces](#)). Parallel programming in these two interfaces are very different. MPI(s) require the software to explicitly send messages between processors. This often leads to very poor performance. Shared memory architectures on the other hand have much lower processor-to-processor latencies than message passing architectures.

Synchronization between processors in one of the most difficult aspects of parallel programming. In messages passing interfaces such as [MPI](#), it is for the programmer to ensure that communication is done scheduled correctly. In [OpenMP](#) such scheduling is not required. But in shared memory interfaces there is the problem of performing several memory operations atomically, which is often required for correct program execution. A lock is a memory location that protects a block of code that is to be run atomically. Only the processor that obtains the lock can execute the atomic block and all others must wait till the lock is released. This had been an inadequate solution to the problem. The current practice is to provide for 'Transactional memory'. It is a hardware mechanism that permits the programmer to define atomic regions (or 'transactions') containing memory accesses to multiple independent addresses. The programmer can define these transactional regions through the instructions provided to the processor ISA. From the programmer's perspective, transactional memory is simply a nice feature that has been added to the shared memory programming interface.

Recent research [C6] shows that OpenMP is easier to program with than MPI. OpenMP with transactional memory is even better than OpenMP with locks. Pointer-based algorithms are difficult to parallelize using MPI. OpenMP also allows significant performance advantage over MPI as the latter uses software based communication. If communications are irregular (and not in large blocks) MPI performance is very poor.

The complexity of the entire process of parallel programming in the Grid context depends on the distribution of processors and also on the extent to which memory is shared between the processors. These aspects are somewhat handled through the resource broker. The main problem in the grid computing context is in the communication overheads involved (which are related to the probability of data loss). This requires some planning based on the type of programs that will be used, the optimal communication network possible and configuration settings.

Parallel compilers are programs that try to parallelize the process of program

compilation. There are again two main approaches to parallel compilation :

## Programming in Existing Languages

The program is coded in an existing language. An optimized compiler is then used to extract parallelism in the program for performance improvement. Naturally a new compiler is required for the extraction process.

## Data Parallel Programming

Based on an extended language specification parallel constructs are first added to the program. This is then converted into standard language by a simple preprocessor. These require complex compilers that can map data-parallel program code into explicit parallel code that correspond to the parallel computer or grid and the available programming tools (libraries, debuggers, etc). The costs involved in maintaining such compilers are justified by the resulting simplifications in developing portable parallel programs and the elimination of the need to explicitly manage concurrency, communication, and synchronization.

## Examples of Parallel Compilers

1  Parafrase Fortran reconstructing compiler: Parafrase is an optimizing compiler preprocessor that takes scientific Fortran code, constructs a program dependency graph and performs a series of optimization steps that creates revised version of the original program and optimize it for high speed architecture.

2. Bulldog Fortran reassembling compiler: The Bulldog compiler is aimed at automatic parallelization at the instruction level. It is designed to catch parallelism not amenable to vectorization. It exploits parallelism within the basic block.

3. Cilk2c compiler: this is a clone compiler which converts <u>CILK</u> source code to C source code. It is bundled with Cilk sources.

4. F90 / <u>High Performance Fortran</u> (HPF) : High performance Fortran extends F90 to support data parallel programming. Compiler directives allow programmer specification of data distribution and alignment. New compiler constructs and intrinsics allow the programmer to do computations and manipulations on data with different distributions.

5. <u>CILK</u> compiler: This is a Cilk compiler developed by the MIT.

6. <u>TopC</u> : A very perfect parallel C.

## **Concurrent Computing**

Concurrent computing is the simultaneous execution of multiple interacting computational tasks one or more processors that may be distributed across a cluster or a grid. The computational tasks may be implemented as a collection of programs, or as a set of processes (or threads) generated by a single program. Concurrent computing differs from parallel computing in that the processors need not necessarily be distributed and in the focus on coordination of tasks. Many aspects of concurrent computing are relevant in the grid computing context.

Concurrent computing systems must be concerned with

- ➢ Proper real-time sequencing of interactions and communications between the different tasks.
- ➢ Coordination of access to resources shared by the tasks

- Robustness of the messaging model and
- Prevention of interference between the concurrent components

Communication between concurrent components of a concurrent computing systems may be masked from the programmer in some implementations (Alice). In others it may be explicit. Again explicit communication may be of the shared memory type or it may be based on message passing. Message passing communication may again be of asynchronous or rendezvous type. Using Java for example it is possible for concurrent components to easily communicate by modifying shared memory locations. Message-passing concurrency is a more robust form of concurrency that is amenable to a wide array of formal techniques (mathematical).

Various models for the analysis of concurrent systems are known. These include
- Petri Nets
- The Actor model,
- Process calculi such as
- π-calculus, Ambient calculus
- Calculus of Communicating Systems
- Communicating Sequential Processes

## Concurrent programming languages

Concurrent programming languages are programming languages that use language constructs for concurrency. These constructs may involve multi-threading, support for distributed computing, message passing, shared resources (including shared memory) or promises (= futures). Most common programming languages provide specific constructs for concurrency. Java for example uses a shared-memory concurrency model. Erlang is one of the most widely used language using a message-passing concurrency model. Almost all existing languages have libraries supporting concurrency to different extents. Some languages that have been specially designed for concurrency are (Multiple compilers are available for each):

- Ada
- Afnix – concurrent access to data is auto-protected.
- Alef – concurrent language with threads and message passing.
- Alice – extension to Standard ML (supports concurrency through futures).
- ChucK – domain specific programming language for audio
- Cilk – a concurrent C developed by the MIT
- C Omega – a research language extending C# (uses asynchronous communication)
- Concurrent Pascal
- Corn
- Curry
- E – uses futures, ensures deadlocks cannot occur
- Erlang – uses asynchronous message passing with nothing shared.
- Join Java – concurrent language based on the Java programming language
- Joule – data flow language that communicates by message passing
- Limbo – relative of Alef, used for systems programming in Inferno (OS)
- Oz – is a multi paradigm language that supports shared-state, message-passing concurrency and futures
- Mozart Programming System – multi platform Oz
- MultiLisp – A parallel implementation of Scheme.
- occam – A language that supports Communicating Sequential Processes (CSP).
- occam-π – A modern variant of OCCAM that involves many parts of Milner's π-calculus
- Pict – essentially an executable implementation of Milner's π-calculus
- SALSA – actor language with token-passing, join and first-class continuations

for distributed computing over the Internet

The installation part of compilers is usually easy. It can be done at nodes of a cluster or as a 'server install' or at the end-user's PC and invoked in the standard way. It is better to keep them available at many more places than is usually required over a Grid. There is no way to do programming without Reading The Fine Manual.

## Some Languages and Compilers:

### Unified Parallel C (UPC)

Unified Parallel C (UPC) is an extension of the C programming language developed in the University of Berkeley and other universities. It is intended for high-performance computing on large-scale parallel machines, including those with a common global address space (SMP and NUMA) and those with distributed memory. The programmer is presented with a single shared, partitioned address space, where variables may be directly read and written by any processor, but each variable is physically associated with a single processor.
UPC extends ISO C 99 with the following constructs for expressing parallelism:
- An explicitly parallel execution model
- A shared address space
- Synchronization primitives and a memory consistency model
- Memory management primitives

The UPC language evolved from experiences with three other earlier languages that proposed parallel extensions to ISO C 99: AC, Split-C, and Parallel C Preprocessor (PCP). UPC is not a superset of these three languages, but rather an attempt to

extract the best characteristics of each. UPC combines the programmability
advantages of the shared memory programming paradigm and the control over data
layout and performance of the message passing programming paradigm.

Most parallel programs are written using message passing with a SPMD model or
shared memory with threads in OpenMP. Global Address Space (GAS) Languages take the
best of both :
- Global address space like threads (programmability)
- SPMD parallelism like MPI (performance)
- local/global distinction (performance)

**Cilk**

Cilk is an algorithmic multi-threaded language. It has been developed at the MIT.
The current version is 5.4.3. Cilk is designed to let programmers concentrate on
structuring their program to expose parallelism and exploit locality and not bother
about scheduling the computations at run-time. The runtime system of the language
takes care of details like load balancing and communication protocols with
guaranteed efficiency and robustness.

The basic Cilk language is very simple. It extends C with three keywords (cilk,
spawn, and sync) to indicate parallelism and synchronization. A Cilk program, when
run on one processor, is semantically equivalent to the C program that results from
the deletion of the three keywords. Such a program is called the serial elision or
C elision of the Cilk program.

Installing the Cilk compiler on GNU/Linux or Unix is simple. One has to just expand
the archive, change directory and follow the generic instructions (./configure &&

make && make install) or indicate any desired configuration options. Any number of extra memory related libraries can be installed.

In order to compile Cilk programs, Cilk 5.4.2.3 (rev 2867) installs the cilkc command, that is actually a special version of the GCC compiler. Files with the ".cilk" extension are taken as Cilk programs by the cilkc command. cilkc accepts many of the same arguments as the gcc compiler.

The Cilk manual is very detailed, generation of parallelism is explained in the following way:" A Cilk procedure may spawn sub-procedures in parallel and synchronize upon their completion. A Cilk procedure definition is identified by the keyword cilk and has an argument list and body just like a C function.
   Most of the work in a Cilk procedure is executed serially, just like C, but parallelism is created when the invocation of a Cilk procedure is immediately preceded by the keyword spawn. A spawn is the parallel analog of a C function call, and like a C function call, when a Cilk procedure is spawned, execution proceeds to the child. Unlike a C function call, however, where the parent is not resumed until after its child returns, in the case of a Cilk spawn, the parent can continue to execute in parallel with the child. Indeed, the parent can continue to spawn  off children, producing a high degree of parallelism. Cilk's scheduler takes the responsibility of scheduling the spawned procedures on the processors of the parallel computer".

## Parallel Java

Manta is a Java system designed for high-performance parallel computing. Manta uses a native compiler and an optimized RMI protocol. The compiler converts Java source code to binary executables. It also deals with serialization and deserialization routines for reducing runtime overheads of RMIs. RMI or Remote Method Invocation is an object-oriented form of the remote procedure call (RPC). RMI is usable for parallel programming too as it cleanly integrates into Java's object oriented programming model.

[Titanium](#) is a Java-based language for high-performance parallel computing. It extends the Java language with immutable classes, fast multidimensional array access, and a parallel SPMD model of communication. The Titanium compiler is an Free and Open Source compiler that translates Titanium into C.

[Spar/Java](#) is a data parallel and task parallel programming language for semiautomatic parallel programming. It is intended for data-parallel applications without support for threads or RMI.

It is difficult to integrate the MPI message-passing style of communication with Java's object-oriented model. The SPMD programming model assumed by MPI is quite different from Java's multi-threading model. Ibis is a parallel Java system developed by Niewenpoort [C10] in particular that avoids this. It is suitable for high-performance computing in grids.

**HPF**

In the [Portland group](#)'s high performance FORTRAN compiler the situation is somewhat like the following. There are many better Free and Open Source versions of HPF compilers including those for GAS versions of HPF and there are some others that are similar in design.

The PGHPF compiler allows for program development in the following way :

- ➢ A new HPF program or a program modified for parallel execution must be

> written with the file having a .hpf, .f90, .for , .F or .f file extension.
> The HPF program must be compiled using PGHPF with the appropriate compiler command line options. This yields an executable binary that links the HPF runtime libraries (default behavior).
> The binary can be executed the target system with scope for using runtime command line options or environment variables.

The PGHPF compiler allows many variations on these general program development steps.

In the SPMD model (Single Processor Multiple Data) each processor is to execute the same program, but must operate on different data. The PGHPF compiler follows this model. This is implemented by loading the same program image into each processor. The processors then operate on their local part of distributed arrays. This is done according to the array sizes, distributions and number of processors as determined at runtime by the compiler or CLI options. This is also dependent on the systems network characteristics. The runtime libraries take into account the communications that are to be performed and are optimized at two levels. At the transport independent level optimal efficient communications are generated based on the data type and the data access pattern used in the computation. The optimization at the transport dependent level involves using a standard communications protocol or a custom data transfer mechanism.

**Conclusion :** Through this exposition we have touched upon some of the main features of modern grid computing that has grown by for and with FOSS.

**Directed References :**

**Important Websites :**

It is necessary to see these for the associated array of documentation too.

**Books / Proceedings:**

**Grid Projects :**

These papers provide nice (but incomplete) overviews of the software available and actual Grid projects :

**Utility Computing :**

## Parallel and Concurrent Programming :

**A. Mani**

**Researcher, University of Calcutta**

**9/1B, Jatin Bagchi Road**

**Kolkata-700029, India**

a_mani_sc_gs@yahoo.co.in

Homepage : http://www.logicamani.in

: